\documentstyle[epsfig]{mn}{}
\begin{document}

\def\mpc{h^{-1} {\rm{Mpc}}}
\def\up{h^{-3} {\rm{Mpc^3}}}
\def\uk{h {\rm{Mpc^{-1}}}}
\def\lsim{\mathrel{\hbox{\rlap{\hbox{\lower4pt\hbox{$\sim$}}}\hbox{$<$}}}}
\def\gsim{\mathrel{\hbox{\rlap{\hbox{\lower4pt\hbox{$\sim$}}}\hbox{$>$}}}}

\title{Properties of groups of galaxies in the vicinity of massive clusters}

\author[C.J. Ragone, M. Merch\'an, H. Muriel \&  A. Zandivarez ]
{C.J. Ragone$^{1,2}$, M. Merch\'an$^{2,3}$, H. Muriel$^{2,3}$ 
\&  A. Zandivarez$^{2,3}$ \\
$^1$Agencia C\'ordoba Ciencia.\\
$^2$Grupo de Investigaciones en Astronom\'{\i}a Te\'orica y Experimental, 
IATE, Observatorio Astron\'omico, Laprida 854, C\'ordoba, Argentina. \\
$^3$Consejo de Investigaciones Cient\'{\i}ficas y T\'ecnicas de la Rep\'ublica 
Argentina. }
\date{\today}

\maketitle 

\begin{abstract}
\label{abs}
This work analyses the  properties of groups of galaxies in the
surroundings of clusters. On the basis of a very large public 
Virgo Consortium  Simulation, we identified systems of galaxies in a wide 
range of masses. Systems with masses greater than 
$ M_{cut}= 4 \times 10^{14}~M_{\odot} ~h^{-1} $
are considered "host", whereas smaller systems are taken as groups.
Our results show that groups properties are affected by the proximity
of massive hosts. Physical properties such as velocity dispersion, 
internal energy ($E$) and virial radius, show an increment, whereas the 
mean density decreases as
the host-group distance is smaller. 
By analysing groups with different properties,
we find that the low mass and the weakly bounded ($E > 0$) 
subsamples, are 
strongly affected by the presence of the host; on the other hand, 
massive groups and groups with $E < 0$ 
do not show dependence on the host-group distance.
Using a sample of groups identified in the final version of the 2dF 
Galaxy Redshift Survey, we find a very similar velocity 
dispersion behaviour in the observational data compared to
results in the simulation. 
We also study the dependence of the groups velocity dispersion on the
host masses in both, observations and simulation; finding that
the larger is the host mass the higher is the effect on its vicinity.
\end{abstract}


\section{Introduction}
\label{int}
In hierarchical models for galaxy formation, structures grow
due to the accretion of smaller systems which formed at earlier times.
According to this scheme, these systems should not only be in the 
cluster itself, but also in its vicinity, harbouring 
an important number of groups or small clusters of galaxies.
In this sense, clusters of galaxies represent high density enhancements 
in the distribution of galaxies. 

On the basis of this scenario, it is natural to ask whether physical 
properties of structures 
depend on its proximity to high density regions, which, as 
previously mentioned, may be well 
represented by rich clusters of galaxies; or, if on the contrary, they
evolve independently on the presence of denser regions. 

Several works have focussed on the evolution of substructures inside
massive systems in simulations or rich galaxy clusters 
(Tormen et al. 1998, Ghigna et al. 2000, Taylor \& Babul 2003, 
De Lucia et al. 2003); 
nevertheless, not much attention has been paid to structures 
in the outer regions of these systems. Hence, it is not well known if 
some of the behaviours of substructures inside the clusters, could also 
be applicable to groups in their periphery.
 
Einasto et al. (2003) have investigated the properties of loose groups in
the Las Campanas Redshift Survey  
(LCLGs), in the vicinity of rich clusters of galaxies, 
including clusters from the Abell 
and APM catalogues, X-ray clusters and also a sample of the richest 
groups from the LCLG catalogue itself. By using an additional sample, 
consisting of all those LCLGs which do not neighbour a rich cluster, they
have been able to compare the properties obtained for dense-environment LCLGs
with those of typical LCLGs. They found that in most cases, the
observed richness of groups near rich clusters is larger than the 
corresponding value for groups in the comparison sample. The same effect
is shown when using Abell counts as a measure of groups richness.
The harmonic radius and the velocity dispersion are also somewhat larger
in the neighbourhood of rich clusters than in typical loose groups. They
also found a strong mass segregation; indicating that loose groups in 
the vicinity of clusters have masses that are larger than mean masses 
of groups drawn from the comparison sample. The same trend is observed 
in the luminosities of LCLGs, showing that groups around rich clusters seem 
to be more luminous. At last, they found that these properties of
loose groups do not show any dependence on clusters richness.

Lemson \& Kauffmann (1999), also address the issue of properties of dark 
matter haloes as 
a function of local density; concluding that mass is the only halo property
which correlates significantly with local environment. In fact, 
high mass haloes are underepresented in low density regions, whereas in high
density regions the situation is the other way round.

In this work we analyse the properties of groups in the vicinity of rich
clusters; taking into account their masses, internal energies, 
velocity dispersions and virial radii. 
To this end, we use data extracted from a high resolution
numerical simulation and from the final version of the 2dF Galaxy Redshift
Survey (2dFGRS) (Colless et al. 2003).
The outline of this paper is as follows. 
In section \ref{simulacion} we describe the simulation that we used and 
give a brief description of the group identification algorithm; 
in section \ref{2dF} we give a description of the 2dFGRS and the identification
of groups in this catalogue; 
section \ref{resultados} contains the results obtained for both, the
simulation (subsection \ref{resulsim}) and the observational data 
(subsection \ref{resul2dF}); 
a summary and a discussion of our results 
are presented in section \ref{conclusiones}.

\section{NUMERICAL SIMULATIONS}
\label{simulacion}
In this study we  use one of the publicly available Very Large 
Simulations (VLS) (Yoshida et al. 2001). 
Given the success of the model to reproduce observational data, we have 
chosen the $\Lambda CDM$ simulation, 
which has a box size of $479~Mpc ~h^{-1}$, 
and $512^3$ particles, resulting in a mass per particle 
$6.86 \times 10^{10}~M_{\odot} ~h^{-1}$.
The cosmological parameters of this simulation are: 
$\Omega_0=0.3$, $\Omega_{\Lambda}=0.7$,
normalisation $\sigma_8=0.9$ and $h=0.7$. Due to the large box size it is 
possible to find a large number of rich systems, which give us the 
possibility to work with a reliable statistical sample. In addition, the
simulation resolution enable small systems to have 
enough number of particles to confidently estimate physical properties.  

The groups identification is carried out using a standard friends-of-friends
algorithm.
The analysis of the VLS haloes catalogue is performed using a
linking length given by $l = 0.17 \bar n ^{-1/3}$, where 
$\bar n$ is the mean numerical density of particles, corresponding to
an overdensity of $\delta\rho/\rho=200$. 
This criterion leads to $\sim 379300$ identified systems with masses ranging
from $1.4 \times 10^{12}~M_{\odot} ~h^{-1}$  (20 particles) to 
$2.35 \times 10^{15}~M_{\odot} ~h^{-1}$.

\section{OBSERVATIONAL DATA}
\label{2dF}
Observational clusters and groups used in this work are extracted from 
the final version of the 2dFGRS, with the best redshifts estimates within
the Northern ($-7^o.5 \leq \delta \leq 2^o.5$, 
$9^h 50^m \leq \alpha \leq 14^h 50^m$), and southern 
($-37^o.5 \leq \delta \leq -22^o.5$, $21^h 40^m \leq \alpha \leq 3^h 30^m$) 
strips of the catalogue. This galaxy sample contains 221414 galaxies.
The group identification was performed using an algorithm similar to that
developed by Huchra \& Geller (1982), modified in order to take into account 
sky coverage problems (Merch\'an \& Zandivarez, 2002), such as 
redshift completeness, which represents a ratio
of the number of galaxies for which redshifts have been obtained to the 
total number of objects contained in the parent catalogue;
and magnitude limit mask, which corresponds to variations of the parent 
survey magnitude limit with the position on the sky.
Also, it was taken into account a minor effect introduced by the magnitude
completeness mask ($\mu$-mask). Mask software was obtained from the
final version data release of the 2dFGRS, kindly provided by Peder Norberg
and Shaun Cole.

In the construction of the group sample, values of $\delta \rho/\rho=80$ 
and $V_0 = 200~km~s^{-1}$ were used to maximise the groups accuracy, 
as stated by Merch\'an \& Zandivarez (2002). 
The group catalogue comprises a total number of 6088 galaxy groups with 
at least 4 members and mean radial velocities in the range 
$900~km~s^{-1} \leq V \leq 75000~km~s^{-1}$.
The minimal number of members is imposed in order to avoid pseudo-groups.

Virial masses were estimated using the virial radius and the velocity
dispersion ($M_{vir}=\sigma^2 R_{vir}/G$, Limber \& Mathews, 1960). For
a better efficiency and stability in the estimation of the velocity 
dispersion, we use the biweight estimator for groups with more than 
15 members, and the gapper estimator
for smaller groups (Beers, Flynn and Gebhardt, 1990; Girardi et al., 1993;
Girardi \& Giuricin, 2000). Finally, the catalogue
mean velocity dispersion is
$253~km~s^{-1}$, the mean virial mass of $7.2 \times 10^{13} \ 
M_{\odot} \ h^{-1}$ and a mean virial radius of $1.04~Mpc ~h^{-1}$.

\section{GROUPS IN THE VICINITY OF CLUSTERS}
\label{resultados}
In order to analyse the behaviour of systems surrounding high density 
regions, we discriminate "host" clusters from "groups" using a 
mass threshold $M_{cut}$.
The studies were carried out in both, simulations and observational data
described in previous sections.

\subsection{Groups properties in the simulation}
\label{resulsim}

\begin{figure}
\epsfxsize=0.5\textwidth 
\hspace*{-0.5cm} \centerline{\epsffile{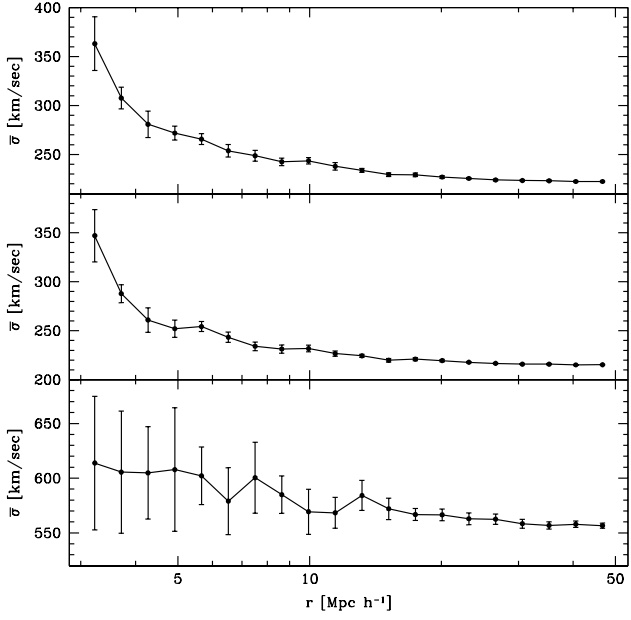}}
\caption
{
Host-groups distances against median velocity dispersion. 
In the top panel the sample corresponds to all groups, in the middle panel
to groups with masses 
$1.4 \times 10^{12}~M_{\odot} ~h^{-1}< M_{group} < 2.0 \times 10^{13}~
M_{\odot} ~h^{-1} $,
and in the lower panel groups masses are 
$2.0 \times 10^{13}~M_{\odot} ~h^{-1} < M_{group} < 4.0 \times 10^{14}~
M_{\odot} ~h^{-1}$. Error bars were 
estimated by the bootstrap resampling technique.
} 
\label{masa}
\end{figure}

Host clusters in the VLS, were selected as systems with masses 
$ M > M_{cut}= 4 \times 10^{14}~M_{\odot} ~h^{-1}$, 
whereas systems with masses 
$1.4 \times 10^{12}~M_{\odot} ~h^{-1}< M < M_{cut} $
are considered groups. 
With this criterion we find $\sim 379000$ groups and 330 hosts, leading to 
a mean host separation of $\sim 70~Mpc ~h^{-1}$.
In this way we have defined our sample of groups neighbouring rich clusters,
enabling us to study the dependence of groups properties on the 
host-group distance. 

The leading aim of our studies, is the dependence of main groups physical
properties on the host distance. Given the asymmetry of these property
distributions, we use the median instead of the standard mean.
We find an increasing trend for the median  
velocity dispersion when considering smaller host-groups distances. Top
panel of figure \ref{masa} shows the above mentioned trend, for all systems
in our group sample. As we are considering  host-group distances starting 
at $3~Mpc ~h^{-1}$, 
miscalculation of group velocity dispersion owing to 
hosts particles contamination, is not likely to happen. Assuring that
the observed effect is not produced by high velocity host particles
misplaced in nearby groups.

\begin{figure}
\epsfxsize=0.5\textwidth 
\hspace*{-0.5cm} \centerline{\epsffile{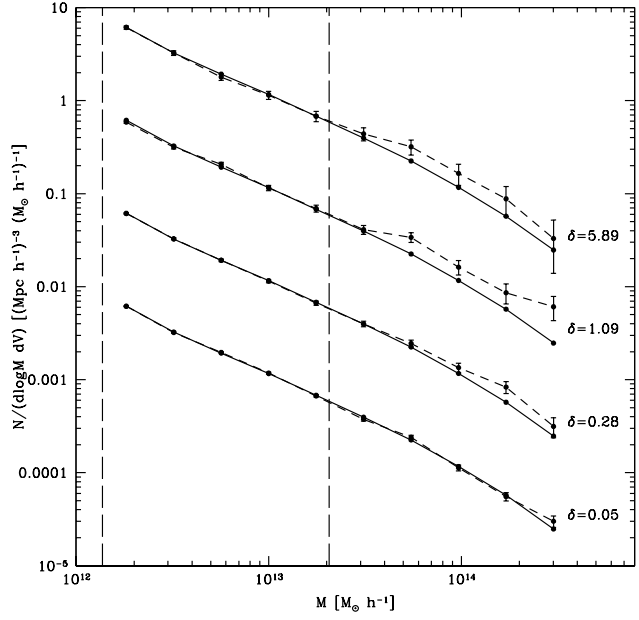}}
\caption
{
Mass function around hosts,
corresponding to different overdensities $\delta$.
The solid line corresponds to the total mass function and dashed lines
to the mass function computed in $1~Mpc~h^{-1}$ thick spherical shells
at radii 3.5, 9.5, 20.5 and 40.5 $Mpc~h^{-1}$, multiplied by $(1+\delta)^{-1}$. 
For clarity, the three first curves and the total one, have been shifted 
(multiplied by 1000,100 and 10 respectively) in the y-axe.
Vertical dashed lines indicate the mass range of the low mass group sample.
Error bars correspond to poisson estimation. 
The total mass function have error bars smaller than the point size.   
} 
\label{fm}
\end{figure}

As stated by Lemson \& Kauffmann (1999), there is a dependence of the 
halo mass function on the environment, which is skewed towards high mass
objects in overdense regions. 
Hence, the rising of the velocity dispersion
in denser environments, could be the result of a higher abundance of 
massive haloes. 
In order to test this possibility, we compute the mass function in
$1~Mpc~h^{-1}$ thick spherical shells, centered on hosts for several radii
(3.5, 9.5, 20.5 and 40.5 $Mpc~h^{-1}$). This allows us to quantify  
the dependence of the mass function on the overdensity 
($\delta$= 5.89, 1.09, 0.28, 0.05 respectively, for the above radii).
Figure \ref{fm} shows in dashed lines the mass function for the mentioned 
radii multiplied by $(1+\delta)^{-1}$ and labelled with the associated 
$\delta 's$; 
solid line, represents the total mass function. Due to the overlapping,
we shift the curves corresponding to the three highest
overdensities by factors 10, 100 and 1000 respectively.
As can be seen, the mass function shape does not change for 
groups with masses 
$1.4 \times 10^{12} M_{\odot} ~h^{-1} < M < 2\times 10^{13} M_{\odot} ~h^{-1}$
(vertical dashed lines in figure \ref{fm}).
Hence, the results obtained from this low mass sample 
will not be affected by the overabundance of massive haloes in high 
density regions.

Based on the previous analysis, we explore the observed velocity dispersion 
behaviour for groups according to their masses; considering
masses greater and lower than $2.0 \times 10^{13}~M_{\odot} ~h^{-1}$.
As a result of this resampling we find that velocity dispersions of 
low mass groups are strongly affected by the presence of host clusters 
(middle panel of figure \ref{masa}), 
whereas high mass groups do not show a significant variation of the median 
when approaching to their associated hosts (lower panel of figure \ref{masa}).  
According to this, we focus the following analysis on this low mass 
group sample. 

\begin{figure}
\epsfxsize=0.5\textwidth 
\hspace*{-0.5cm} \centerline{\epsffile{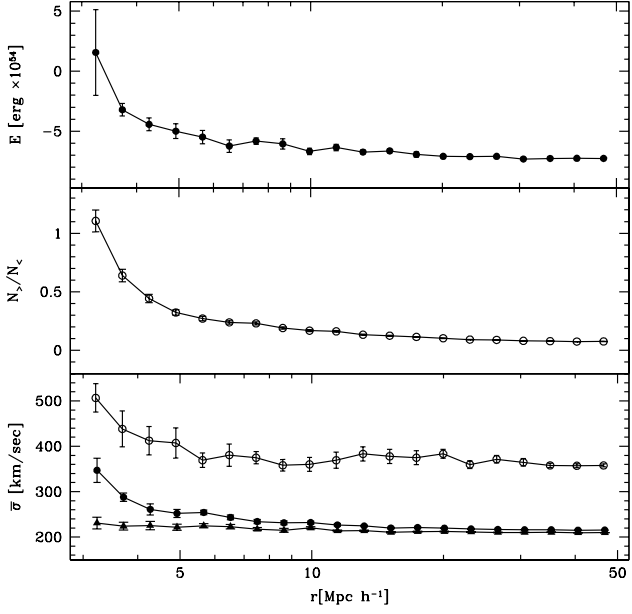}}
\caption
{
Lower panel shows host-groups distances against the median velocity
dispersion. 
Dots: low masses group sample. Open circles: groups with energy 
$E > 0$.
Triangles: groups with energy $E < 0$. Error bars were estimated
by the bootstrap resampling technique.
In the middle panel is plotted the ratio between the
number of groups with $E > 0$ 
and $E < 0$, as a function of the host-group distance.
In this case error bars correspond to error propagation technique.
In the top panel is shown the median energy against host-group distance. 
Bootstrap resampling technique was used to obtain the error bars.
} 
\label{sigE2}
\end{figure}

At first sight, the rising of the median velocity dispersion, could be
a result of the gravitational influence of the host on its neighbourhood. This
possibility encourages us to analyse the behaviour of the internal energy, 
$E$, in the same way we did for the velocity dispersion.
This quantity is estimated by:
$$ E=\frac{1}{2}\sigma^2 M_{group} - \frac{G m^2 N (N-1)}{R_{vir}}$$
where the first and second terms correspond to the kinetic and potential
energies respectively,
$\sigma$ is the 3D velocity dispersion, $M_{group}$ the group mass,
m the particle mass, N the number of members of a group and $R_{vir}$ 
the group virial radius. 
Top panel of figure \ref{sigE2} shows the dependence
of the group median energy on the host distance. It is clear 
that groups situated closer to high density regions, tend to have
higher energies, indicating that these groups are typically weakly bounded.
Next, we subdivide the low mass group subsample into
positive and negative internal energy groups. In this case, 
we find that the rising median velocity dispersion trend, is maintained for 
positive energies groups (open circles in bottom panel of figure \ref{sigE2}), 
whereas the negative energies ones, show a constant behaviour (filled triangles
in bottom panel of figure \ref{sigE2}). 
Moreover, positive
energy groups not only are the main contributors to the signal, 
but also dominate in number
towards central regions; as can be seen in the middle panel of the same
figure. 
This result suggests that the rising of the median velocity
dispersion towards smaller host-group distances, has to do with an
equilibrium loss, due to an increment in the internal velocities of 
these groups.
  
\begin{figure}
\epsfxsize=0.5\textwidth 
\hspace*{-0.5cm} \centerline{\epsffile{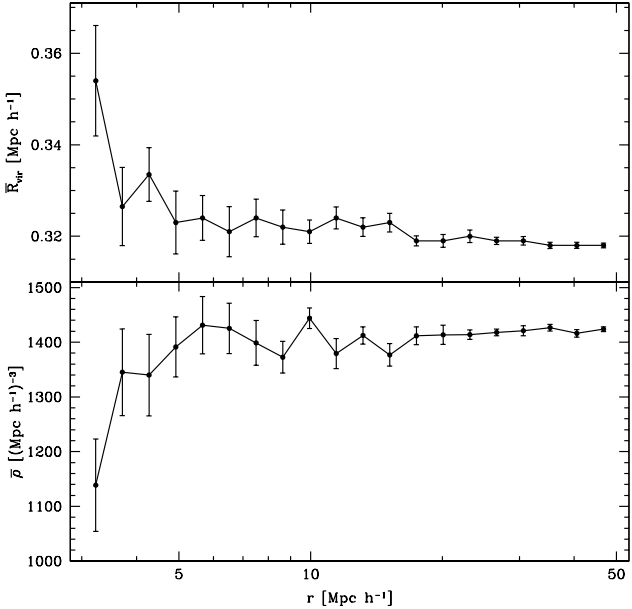}}
\caption
{
Median virial radii (Top panel) and 
median mean density (Bottom panel) against host-group distance. 
Error bars are estimated using the bootstrap resampling technique.
} 
\label{prop}
\end{figure}

To complement the previously obtained results, we have also analysed  
the dependence
of the median virial radius on the host distance. As shown in the top panel
of figure \ref{prop}, the median virial radius weakly increases towards inner 
regions. 
This is in agreement with the previous analysis, since contributes
to a decrease of the potential energy with the consequent increment in the 
internal energy. 
In the bottom panel of figure \ref{prop}, the median density of groups,
$\bar\rho =\overline{( N/R_{vir}^3)}$, is plotted against 
the host-group distance showing that groups nearer to hosts, 
have lower mean densities than groups in the outer regions. 

\begin{figure}
\epsfxsize=0.5\textwidth 
\hspace*{-0.5cm} \centerline{\epsffile{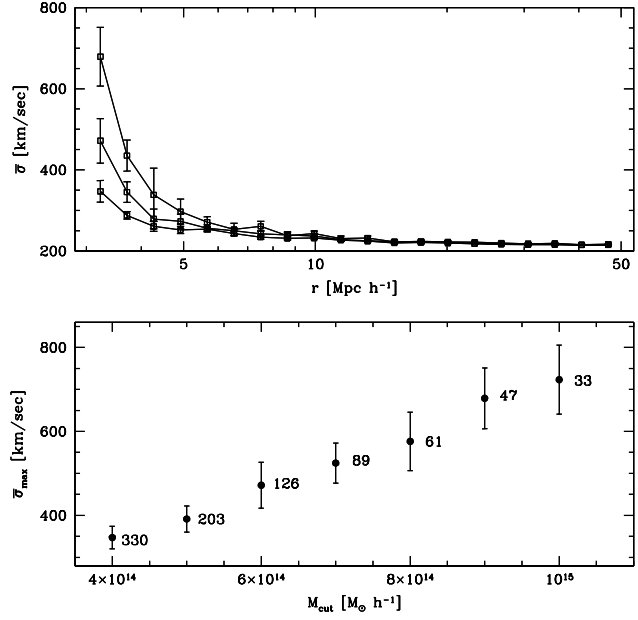}}
\caption
{
Upper panel: median velocity dispersions for 
$M_{cut} = 4 \times 10^{14}~M_{\odot} ~h^{-1}, 
6 \times 10^{14}~M_{\odot} ~h^{-1} $ and $9 \times 10^{14}~M_{\odot} ~h^{-1} $ 
corresponding to the bottom, middle and upper trend respectively.
Bottom panel: median velocity dispersion in the nearest bin of 
the host-group distances ($\bar{\sigma}_{max}$), against $M_{cut}$.
Numbers beside points indicate the number of host with masses larger 
than $M_{cut}$.
Error bars are estimated according to the bootstrap resampling technique.
} 
\label{host}
\end{figure}

A possible interpretation of the above results can be found analysing
the dynamical properties of systems of galaxies in the presence of
external gravitational fields. The interaction
between two systems of very different size will mainly  affect the
properties of the smaller one.  
The tidal radius $R_t$ specifies the size of the smaller 
system beyond 
which the mass becomes bound to the potential of the larger one. For a
group with total mass $M_g$ in the neighbourhood of a host with mass $M_h$,
the tidal radius can be roughly approximated by
$$R_t \simeq r \left(\frac{M_g}{3M_h}\right)^{\frac{1}{3}}$$
where $r$ is the host-group distance (Binney \& Tremaine, 1987). For our
low mass group sample, the mean group mass is 
$\sim 4.0 \times 10^{12}~M_{\odot}~h^{-1}$; whereas host sample has a 
mean mass of $\sim 6.5 \times 10^{14}~M_{\odot}~h^{-1}$. 
Replacing these values in the above equation, for a distance 
of $3~Mpc~h^{-1}$, a value of $R_t = 0.38~Mpc~h^{-1}$ is obtained, which
is quite similar to the median virial radius at the same distance 
(top panel of figure \ref{prop}). 
This result shows that in the proximity of large systems, tidal forces
exerted by the host group can be as strong as groups internal forces.

In an attempt to explore the dependence of groups properties 
on the hosts masses, we compute median velocity 
dispersions in the same way we did before, for different mass thresholds.
In this case we leave a gap between the upper mass limit used to select groups, 
and the mass threshold used to select hosts, in order to work with the same
low mass group sample we did before.
In top panel of figure \ref{host}, we show the median velocity dispersions
against host-group distance, for three different host samples using 
$M_{cut} = 4 \times 10^{14}, 6 \times 10^{14}$ and $9 \times 10^{14}~
M_{\odot} ~h^{-1}$, corresponding to the lower, middle and upper
trend respectively; the more massive are the selected hosts, the
steeper is the median velocity dispersion trend. 
In the bottom panel of figure \ref{host} we plot the value of the maximum median velocity dispersion,
$\sigma_{max}$, corresponding to groups in the distance range from 
$3.0$ to $3.45~Mpc ~h^{-1}$, as a function of $M_{cut}$. 
A strong enhancement of the effect that hosts exert on its neighbourhood,
is observed when considering more massive hosts. 
This fact supports the 
scenario where the gravitational potential of massive systems,
perturbs the behaviour of the surrounding smaller groups.

\subsection{Groups properties in the 2dFGGC}
\label{resul2dF}

In this section we describe the results obtained for the sample of
groups derived from the 2dFGRS as described in section \ref{2dF}.

\begin{figure}
\epsfxsize=0.5\textwidth 
\hspace*{-0.5cm} \centerline{\epsffile{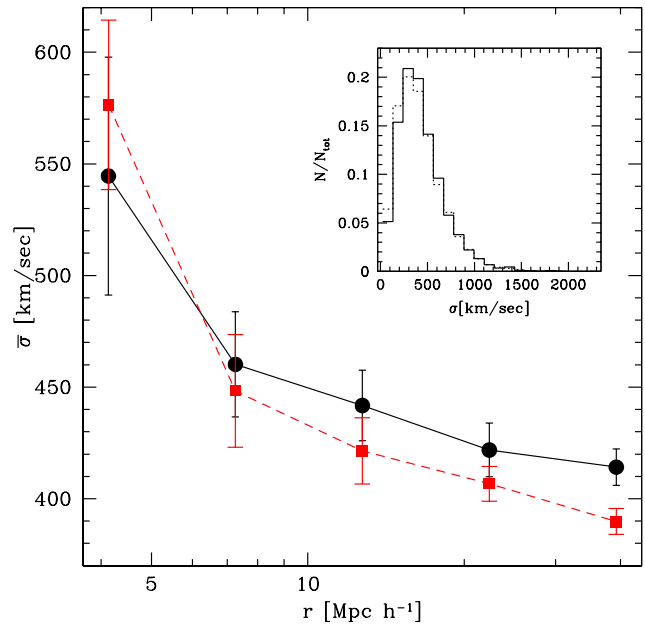}}
\caption
{
In the inset panel, the solid line represents the 2dFGGC normalised 
velocity dispersion distribution; the dashed line is for a subsample of
VLS groups which imitates the former.
Velocity dispersion medians for both samples, are plotted in the
main panel for 
$M_{cut}=7 \times 10^{14}~M_{\odot} ~h^{-1}$,
the line types correspond to the same 
samples as for the inset panel. Bootstrap resampling technique is used 
for the computation of error bars.
} 
\label{mediana}
\end{figure}

\begin{figure}
\epsfxsize=0.5\textwidth 
\hspace*{-0.5cm} \centerline{\epsffile{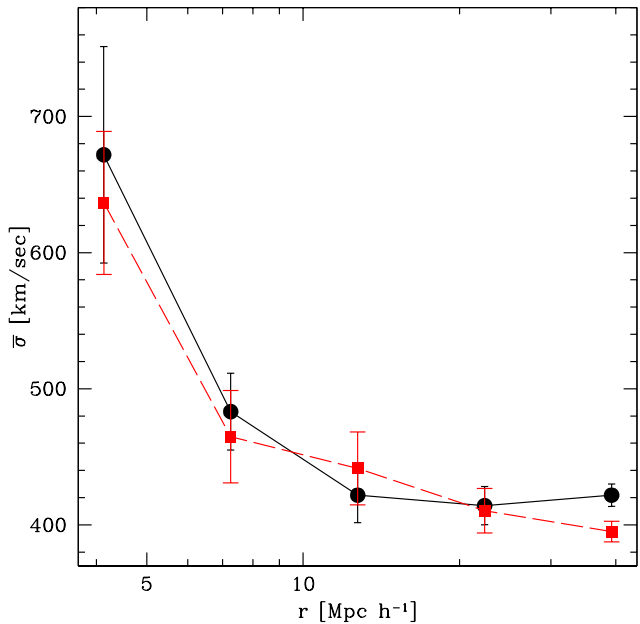}}
\caption
{
Median velocity dispersion for 
$M_{cut}=9 \times 10^{14}~M_{\odot} ~h^{-1}$
Line types correspond to the same samples as for figure \ref{mediana}.
Bootstrap resampling technique is used 
for the computation of error bars.
} 
\label{mediana9}
\end{figure}

In order to compare with the results of the simulation, we extract a 
random subsample from
groups in the VLS, that reproduces the 2dF group velocity dispersion 
distribution. 
Inset plot of figure \ref{mediana}, shows in dashed line the normalised
distribution imposed for VLS groups, and in solid line the 
corresponding to 2dF groups.
We use for both samples $M_{cut}=7 \times 10^{14}~M_{\odot} ~h^{-1}$; 
this choice leads to
45 hosts and 4548 groups in the 2dF, and 156 hosts and
38034 groups in the VLS.
We compute, as in previous section, the median velocity dispersion of groups;
in figure \ref{mediana} solid line represents the median velocity dispersion
for 2dF groups, whereas dashed line is associated to VLS groups. 
This figure shows an
enhancement in the velocity dispersion, as predicted in the previous section 
using the total VLS group sample.
The mentioned behaviour, is consistent with Einasto et al. (2003), 
since they state, that the velocity dispersion of loose groups
in the neighbourhood of rich clusters is about 1.3 times larger than that in
field groups; value which is very similar to the ratio between the 
first and last
point of figure \ref{mediana}. 
They also find that groups virial masses in high
density environments are about 2.5 times larger than masses of field groups,
which is also in agreement with our results, since the virial radii
for 2dF groups show a very weak increment towards smaller host-group 
distances, and the virial mass is $M_{vir} \propto R_{vir} \sigma^{2}$. This
results in an evident increment on the groups virial masses
towards the hosts, mainly due to the velocity dispersion behaviour.
On the other hand, we have seen in section 4.1, that
real masses, do not have a dependence on
the host-groups distance; indicating that virial masses might not
be a reliable indicator of the real mass of a group nearby to a high
density region.
In order to study the dependence of groups velocity dispersion on 
the host masses for the 2dF sample, we take a higher mass threshold,
$M_{cut}=9 \times 10^{14}~M_{\odot} ~h^{-1}$, resulting in 25 hosts,
and perform the same kind of analysis. 
Figure \ref{mediana9} shows a higher signal for the velocity dispersion
in the inner regions which is in concordance with the results we found
for the simulation; the larger is the host mass, the higher is the 
typical velocity dispersion of groups near to the host.

\section{CONCLUSIONS}
\label{conclusiones}

Given the large volume and high resolution of the  publicly available 
VLS, we could identify a large number of systems of galaxies in a wide 
range of masses. 
Using a mass threshold we defined several samples of host 
systems and groups.
 
Our results show that the median velocity dispersion of groups of 
galaxies increases
as the distance to the host clusters is smaller. 
Taking subsamples of groups, considering high and low masses,
we find that the  previous behaviour is not present in the high mass subsample
whereas low mass groups account for the observed effect. 
This low mass subsample was used to study
the behaviour of groups according to their internal energies. 
We found that the internal energy of groups increases towards smaller values of 
host-group distances, showing up that the gravitational equilibrium of groups
is influenced  by the presence of massive systems. 
By splitting this sample in two groups subsamples with positive and negative 
energies, we found that only those with positive energies increase 
the median velocity dispersion towards smaller host distances.
In order to unmask the source of these effects, we studied the dependence
on the distance of complementary physical properties. By considering smaller
host-groups distances, the virial radius increases and the opposite trend
is observed for the mean density. 

A possible scenario describing the above results, may be placed within the 
framework of the tidal forces 
exerted by the  massive clusters over neighbouring systems. 
In the proximity of a cluster, its potential has a strength which is similar 
to the one exerted by the group itself; indicating that tidal forces are
likely to affect groups properties.

Finally, analysing different  subsamples  
of host clusters we found that, the more massive is the host cluster 
the greater is the effect on the group velocity dispersion,
which is also consistent with the previous model. 

Using the final version of the 2dF redshift survey, we repeated the velocity
dispersion analysis performed to the VLS groups. 
In this case we take a subsample of VLS groups, which reproduces the velocity
dispersion distribution of groups identified in the 2dFGRS. The trend we find 
for the observational data is in excellent agreement with the result we
obtain for the simulation subsample. Moreover, 2dF groups velocity dispersion 
show a dependence on the host masses very similar to that found in 
the simulation.
The good agreement  between the results coming from the 
numerical simulation and the observational data suggests that no astrophysical
mechanisms, other than the  gravitational forces, are needed in order to 
explain the dynamical properties of groups of galaxies.

\section*{Acknowledgements}
We thank the referee for useful suggestions that improved the original version
of the paper.
The simulations in this paper were carried out by the Virgo Supercomputing 
Consortium using computers based at Computing Centre of the Max-Planck 
Society in Garching and at the Edinburgh Parallel Computing Centre. 
The data are publicly available at www.mpa-garching.mpg.de/NumCos.
We thank to Peder Norberg and Shaun Cole for kindly providing the software
describing the masks of the 2dFGRS and to the 2dFGRS Team for having made
available the final version of the catalogue.
This work has been partially supported by Agencia Nacional de Promoci\'on
Cient\'{\i}fica y T\'ecnica,
Secretar\'{\i}a de Ciencia y T\'ecnica (SeCyT), 
the Agencia C\'ordoba Ciencia and Fundaci\'on Antorchas, Argentina.

\end{document}